# Devil´s staircase like behavior of the range of random time series associated with the tangled nature of evolution

Alexander S. Balankin,[1,2] Oswaldo Morales M.

[1)] Sección de Posgrado e Investigación, ESIME, Instituto Politécnico Nacional,

México D.F., México 07738

[2)] Grupo "Mecánica Fractal", México, http://www.mfractal.esimez.ipn.mx

We present empirical evidence that the range of random time series associated with the tangled nature model of evolution exhibits a devil's staircase like behavior characterized by logarithmic trend and the "universal" multi-affine spectrum of scaling exponents $\zeta_q$ of $q \leq q_C$ moments of q-order height-height correlations, whereas for $q > q_C$ the moments behaves logarithmically.

89.75.Da, 89.65.Gh, 05.40.-a, 05.45.Tp

Many complex systems evolve through periods of relative quiescence separated by brief outbursts of hectic activity.[1] The intermittent activity has been observed in a great variety of systems studied in physics,[2] biology,[3] geosciences,[4] and econophysics.[5] While the punctuated equilibrium behavior frequently associated with avalanche dynamics outlined for Self-Organized Critical systems [[16]], recently it was pointed out[7] that the similar behavior also is also characteristic of Tangled Nature model of evolution.[8] Accordingly, to model a system displaying punctuated dynamics, first of all we need to distinguish between two fundamentally different scenarios of system evolution.

The Self-Organized Criticality (SOC) focus on the spatio-temporal dynamics of avalanches which build up long-range correlations in the system,[9] whereas the Tangled Nature (TN) model[10] stresses the increase stability of sequential metastable states, leading to a slowing down of the pace of evolution [[7,8]]. Time series analysis with methods from statistical physics allows us to develop and verify macroscopic models of complex systems evolution on the basis of data analysis.[11] Specifically, the analysis of the scaling properties of the time series fluctuations has been shown to give important information regarding the underlying processes responsible for the observed macroscopic behavior [[10]]. Accordingly, time series of appropriate observables, $p(t)$, can be analyzed to distinguish between different scenarios leading to punctuated equilibrium behavior.

The essential feature of TN model is the logarithmic slow-down of the evolution [[7]] in contrast to scale-invariant dynamics associated with SOC [[1,6,9]]. In [[7]] the time record $M(t) = \max p(t)$ is used to illustrate the difference between SOC and TN evolution. It was found that the correlations between the consecutive quakes (changes of $M(t)$) are negligible and the consecutive quake waiting times are statistically independent. In this Letter, we show that the scaling analysis of time series range records, $R(t) = \max p(t) - \min p(t)$, permits to distinguish between SOC and TN scenarios of system evolution and gives additional information about the correlations in system dynamics.

The fluctuations of any time series can be characterized by the magnitude (absolute value) of changes and their direction (sign).[12] The magnitude series relates to the nonlinear properties of the original time series, while the sign series relates to the linear properties [11]. It was found that the magnitude of fluctuations of many apparently random time series exhibits fat-tailed power-law distribution and display long range power law correlations, characterized by the so-called Hurst exponent $\zeta$ [10,11]. The sign time series also exhibit the scale-invariant dynamics but with different scaling exponent $\zeta_{sign}$ [11]. Moreover, the scaling properties of negative and positive changes of real-world time series may be different.[13][14] This asymmetry should be reflected in the scaling behavior of time series range $R(t)$.

In this work, we analyzed the historical price records (in constant US dollars) of some commodities[15] (crude oil, natural gas, gold). Early, the fluctuations in these time series were studied in Refs.[16][17][18] It was found that the magnitude of price changes, $|\Delta P(t,\tau)|$ exhibit long-range power-law correlations, nevertheless the price $P(t)$ and the price changes $\Delta P(t,\tau) = P(t+\tau) - P(t)$ are uncorrelated beyond rather short time scales [14,16]. The distributions of negative and positive changes are fat-tailed and characterized by slightly different exponents [14]. We note that these properties are consistent with the SOC, as well as with the TN scenarios of market evolution.[19]

Accordingly, in this work we focus on the scaling behavior of $R(t)$ and its relation to the scaling behavior of the sign and magnitude of price changes. To test the correlations in the analyzed time series we studied the autocorrelation function $C(\tau) = \langle p(t+\tau)p(t)\rangle_T / \langle p^2(t)\rangle_T$, where the angle brackets denote the time average. The scaling properties of time series and their ranges were studied by calculating the q-order height difference correlation function

$$\sigma_q(\tau) = \left\langle |p(t) - p(t+\tau)|^q \right\rangle_T^{1/q} \propto \tau^{\zeta_q}, \qquad (1)$$

for $0.01 < q \leq 100$; where $\zeta_q$ is the spectrum of scaling exponents.[20] Furthermore, the Hurst exponent, $\zeta = \zeta_2$, of each time series was also determined from the scaling behavior of power spectrum, $S(\omega)$, and rescaled range, $R/S$. Specifically, we explored the following scaling behavior: $S(\omega) = \langle \hat{p}(\omega)\hat{p}(-\omega)\rangle \propto \omega^{-(2\zeta+1)}$, where $\hat{p}(\omega) = T^{-1/2}\int dx [p(t) - \langle p(t)\rangle_T] \exp(i\omega t)$ is the Fourier transform of $p(t)$; $V(\tau) = \left\langle [p(t) - p(t+\tau)]^2 \right\rangle_T \propto \tau^{2\zeta}$; and $R/S \propto \tau^{\zeta}$, where $R/S$ defined as the ratio of the maximal range of the integrated time series to its standard deviation (see also Ref.[21]).

Figure 1 (a,b) shows the daily records of the spot prices $P(t)$ and price changes $\Delta P(t) = P(t+1) - P(t)$ from the West Texas Intermediate crude oil price listings [[15]]. To avoid the effect of inflation we analyse the crude price in constant 2003 US dollars over the period from 2 January 1986 to 28 May 2004 representing 4652 observations (weekends and business holidays are excluded). We find (see Fig. 2 a) that the autocorrelation function of price record decays exponentially as $C \propto \exp(-\tau/\tau_0)$ with a characteristic time $\tau_0 = 120$ business days (about the half business year). Furthermore we find that $\zeta_q = \zeta_2 = \zeta = 0.5 \pm 0.02$ (see Fig. 2 (c-d)). So, there are no long-range correlations in the crude oil price record. This is consistent with the finding that the crude oil spot price distribution is a symmetric logistic distribution (see insert in Fig. 1 (a)).

At the same time, we note that the absolute values negative changes, $|\Delta P_-|$, generally are larger then positive changes, $\Delta P_+ > 0$ (see Fig. 1 (b)), while the number (frequency) of positive changers $N_+(t)$ are slightly larger then the number of negative changes $N_-(t)$. We find that the difference $\Delta N(t) = N_+ - N_-$ possesses linear trend (see Fig. 1 (c)), whereas the deference between absolute values of consecutive ordered negative and positive changes scales as $\Delta(n) = |\Delta P_-| - \Delta P_+ \propto n^{-0.3}$ (see insert in Fig 1 (b)). As the result of these "leverage effects", the price range $R(t)$ displays stepwise increase with logarithmic trend (Fig. 1 (d)) expected in TN model of market evolution.[22]

Furthermore, we find that rang increments (Fig. 3 (a)) are distributed according to the fat-tailed log-logistic distribution (Fig. 3 (b)) with Lévy index $\mu = 2.58$ out of Lévy stable range ($0 < \mu < 2$). This indicates the presence of long-range correlations in the price range behavior. Scaling analysis shows (see Fig. 3 (c-e) that $R(t)$ has the devil's staircase like behavior (Fig. 1 (d)), characterized by the "universal" spectrum of scaling exponents (see insert in Fig. 3 (f)):

$$\zeta_q = \zeta_*(1+\frac{\alpha}{q}), \text{ where } \alpha = \mu - 1, \text{ for } 0.1 \leq q_C, \qquad (2)$$

with $\zeta_* = 0.31 \pm 0.01$ and $\alpha = 1.58 \pm 0.02$ ($\zeta_2 = 0.56 \pm 0.02$, see Fig. 3 (c)); *i.e.*, $R(t)$ displays persistence. At the same time, we found that the moments with $q > q_C = 2.15$ depend logarithmically on $\tau$, e.g.

$$\sigma_q(\tau) = b(q)\ln\tau - a(q),  \qquad (3)$$

where $a(q)$ and $b(q)$ are decreasing functions of $q$. The critical value $q_C$ is defined as $R^2(1) \geq R^2(3)$ for $q \leq q_C$ and $R^2(1) < R^2(3)$ when $q > q_C$, (see Fig. 3 e). The transition from power-law (1) to logarithmic behavior (3) of the q-order height difference correlation function is consistent with the logarithmic trend of $R(t)$.

It should be emphasized that the same results also were obtained for all price records of length 3650 observations between different dates within the original limits (from 2 January 1986 to 28 May 2004). Further, we find that the ranges of all studied price commodities display the devil's staircase like behavior with logarithmic trend, characterized by spectrum of scaling exponents (1) with $0.25 < \zeta_* \leq 0.5$, and $1 < \alpha < 2$; and $2 < q_C < 3$. Accordingly, the Lévy index, $\mu = \alpha + 1$, for fat-tiled distributions of range increments of all studied price commodity records is found to be out of Lévy stable range. Detailed results of these studies will be published elsewhere.

This work was supported by the Mexican Government under the CONACyT Grant No. 44722 and the National Polytechnic Institute under research program N 346.

[19] The commodity markets is defined in near 2 dimensions, one dimension is time and another represents ensemble of stocks.

**Figure 1.** (a) Time records of West Texas Intermediate crude oil spot price in the current (1) and in the 2003 constant (2) dollars per barrel, $/bbl (source: Bloomberg database [15]), and the moving average of price in constant dollars [the insert shows the conditional probability distribution of prices in constant dollars]. (b) Time record of price changes [insert: $\Delta n$ vs. $n$ in the log-log coordinates]. (c) The graph of $\Delta N$ vs. the calendar time. (d) Time record of price range [the insert shows the logarithmic trend of $R(t)$].

**Figure 2.** (a) Autocorrelation function of price record (2) shown in Fig 1(a) in semilog coordinates [insert shows the graph $C(\tau)$]. (b)-(d) Fractal graphs of price record obtained by (b) the rescaled-range and (c) the power-spectrum methods; and (c) the q-order height difference correlation analysis (from bottom to top: $q = 0.1, 0.2, 0.5, 1, 2, 3, 5$).

**Figure 3.** (a) Time record of price range changes. (b) Conditional probability distribution of $\Delta R$ [insert: the distribution trend in the log-log coordinates]. (c) Power spectrum of the price range record shown in Fig. 1(d). (d) and (e) Graphs of $\sigma_q(\tau)$ in the log-log coordinates: (d) from bottom to top $q = 0.01, 0.015, 0.025, 0.05, 0.1$; (e) $q = 0.5$ (1), $q = 1$ (2), and $q = 2.2$ (3) [solid lines – the power law fits: (1) $\sigma_{0.5} = 0.0001\tau^{1.234}$ $R^2 = 0.9995$; (2) $\sigma_1 = 0.017\tau^{0.745}$, $R^2 = 0.9969$; (3) $\sigma_2 = 0.2741\tau^{0.492}$, $R^2 = 0.9695$; and pointed lines – the logarithmic fits: (1) $\sigma_{0.5} = 0.02\ln\tau - 0.056$, $R^2 = 0.8741$; (2) $\sigma_1 = 0.214\ln\tau - 0.474$, $R^2 = 0.9689$; (3) $\sigma_2 = 0.787\ln\tau - 1.129$, $R^2 = 0.9986$. (f) Graph

of $\zeta_q$ vs. $1/q$ (dots – experimental data, solid line – data fit by eq. (2) for $0.01 \leq q \leq 2.15$); the insert shows the graph of $\zeta_q$ vs. $q$.

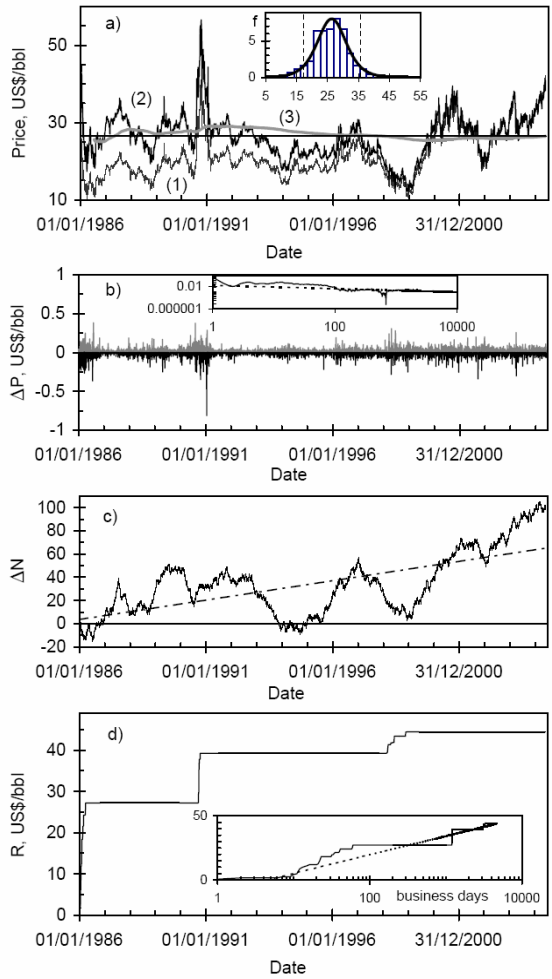

Figure 1.

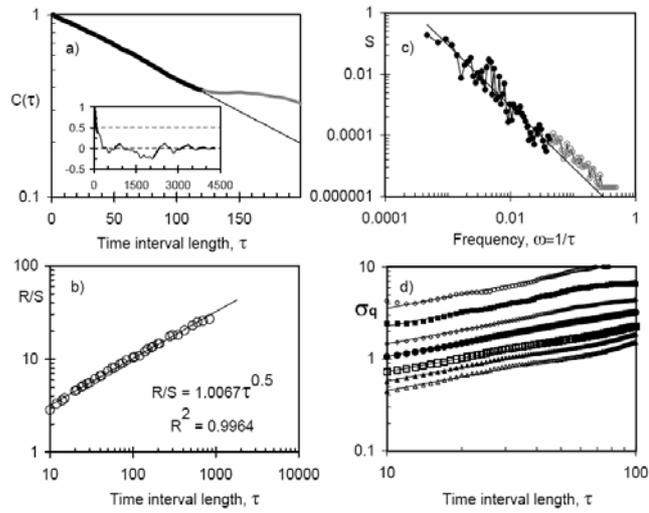

Figure 2

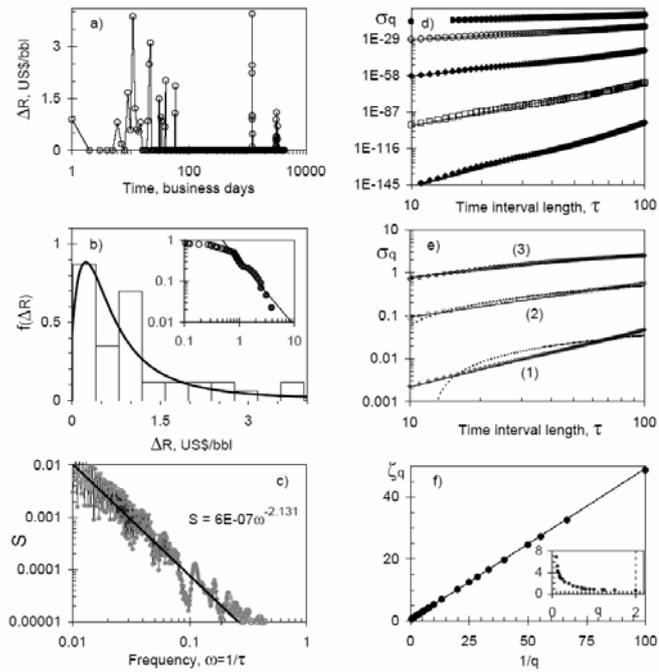

Figure 3.